\DocumentMetadata{}
\documentclass[sigconf,review]{acmart}

\makeatletter
\def\mdseries@tt{m}
\makeatother

\pdfoutput=1

\usepackage{microtype}
\usepackage{graphicx}
\usepackage{subcaption}
\usepackage{booktabs}
\usepackage{comment}
\usepackage[frozencache=true,cachedir=minted-cache]{minted}
\usepackage{amsmath}
\usepackage{mathtools}
\usepackage{amsthm}
\usepackage[capitalize,noabbrev]{cleveref}
\usepackage[textsize=tiny]{todonotes}
\usepackage[T1]{fontenc}
\usepackage{beramono}
\usepackage{pgfplots}
\usepackage{pgfplotstable}
\usepgfplotslibrary{statistics}
\usepgfplotslibrary{groupplots}
\pgfplotsset{compat=1.18}

\usepackage{tikz}
    \usetikzlibrary{positioning, arrows, automata, decorations.pathreplacing}
\usepackage{listings}
\usepackage{algorithm}
\usepackage{algorithmic}
\usepackage{multirow}

\definecolor{dkgreen}{rgb}{0,0.6,0}
\definecolor{gray}{rgb}{0.5,0.5,0.5}
\definecolor{mauve}{rgb}{0.58,0,0.82}

\lstdefinelanguage{docker}{
  keywords={FROM, RUN, CMD, LABEL, EXPOSE, ENV, ADD, COPY, ENTRYPOINT, VOLUME, USER, WORKDIR, ARG, ONBUILD, STOPSIGNAL, HEALTHCHECK, SHELL},
  keywordstyle=\color{blue}\bfseries,
  identifierstyle=\color{black},
  sensitive=false,
  comment=[l]{\#},
  commentstyle=\color{purple}\ttfamily,
  stringstyle=\color{red}\ttfamily,
  morestring=[b]',
  morestring=[b]"
}

\newcommand{\approachName}{PerfBench}

\begin{document}
\title{PerfBench: Can Agents Resolve Real-World Performance Bugs?}

\author{Spandan Garg}
\authornote{Corresponding author.}
\affiliation{%
  \institution{Microsoft Corporation}
  \streetaddress{One Microsoft Way}
  \city{Redmond}
  \state{WA}
  \country{USA}}
\email{spgarg@microsoft.com}

\author{Roshanak Zilouchian Moghaddam}
\authornote{Work done while at Microsoft.}
\affiliation{%
  \institution{Microsoft Corporation}
  \streetaddress{One Microsoft Way}
  \city{Redmond}
  \state{WA}
  \country{USA}}
\email{rozilouc@microsoft.com}

\author{Neel Sundaresan}\authornotemark[2]
\affiliation{%
  \institution{Microsoft Corporation}
  \streetaddress{One Microsoft Way}
  \city{Redmond}
  \state{WA}
  \country{USA}}
\email{neels@microsoft.com}

\date{}

\begin{abstract}
Performance bugs are inefficiencies in software that waste computational resources without causing functional failures, making them particularly challenging to detect and fix. While recent advances in Software Engineering agents have shown promise in automated bug fixing, existing benchmarks primarily focus on functional correctness and fail to evaluate agents' abilities to identify and resolve non-functional issues like performance bugs. We introduce \approachName{}, a benchmark comprising 81 real-world performance bug-fixing tasks from popular .NET repositories on GitHub. Unlike existing benchmarks that rely on pre-existing test suites, \approachName{} features a novel evaluation harness that allows agents to generate their own performance benchmarks and validates fixes by comparing execution metrics collected for developer fix and agent fix. Each task in \approachName{} is derived from actual developer fixes linked to performance-related issues, which are then verified by human experts, ensuring real-world relevance. Our evaluation reveals that current state-of-the-art coding agents struggle with performance optimization tasks, with baseline OpenHands agent achieving only a $\sim$3\% success rate on our benchmark. We develop OpenHands-Perf-Agent, which incorporates performance-aware tooling and instructions and achieves a $\sim$20\% success rate on the benchmark. We show that by ensuring the agent has proper instructions to benchmark its changes and tooling for benchmark output processing, we can improve the agent performance significantly, but room for improvement still remains. \approachName{} provides a challenging test set for furthering the capabilities of agents in fixing perf issues.
\end{abstract}

\maketitle

\thispagestyle{empty}
\pagestyle{plain}

\section{Introduction}
\label{sec:introduction}

Performance bugs represent a unique class of software defects that impact the application's efficiency without causing any functional failures. Unlike traditional bugs that manifest as crashes or incorrect outputs, performance bugs silently waste computational resources, increase latency as well as costs~\cite{attariyan2012x, dean2014perfscope}. These inefficiencies are particularly problematic in cloud applications where resource consumption directly translates to computational costs, and can also impact end-user experience due to problems such as increased latency.

The rise of Software Engineering agents has revolutionized automated software engineering, with proprietary systems like Claude Code~\cite{claude_code_2024}, Copilot Agent~\cite{copilot_agent_2024}, Windsurf~\cite{windsurf_2024}, Devin~\cite{cognition2024devin}, etc. as well as open-source agents such as SWE-agent~\cite{yang2024sweagent}, and OpenHands~\cite{wang2024opendevinopenplatformai} demonstrating impressive capabilities in fixing functional bugs as well as other software engineering tasks such as test generation, code search, feature development, etc. However, current bug fixing benchmarks for evaluating these agents, such as SWE-bench~\cite{jimenez2024swebench}, focus exclusively on functional correctness. This leaves a significant gap in understanding how well these agents can handle non-functional bugs such as performance or security bugs.

This gap exists for several reasons. Firstly, we argue that performance bug fixing requires a fundamentally different high-level sequence of steps compared to the ones needed to fix functional bugs. The agents need to understand how humans expect performance bugs to be fixed and what kind of constraints need to be met for the changes to considered acceptable by developers, such as unit test success, as well as performance improvement and no performance deterioration in other aspects of the codebase. Agents must also understand performance-related concepts like computational complexity, resource utilization, and performance trade-offs between resources. Validating performance improvements also requires infrastructure for benchmarking and comparison between benchmarks, unlike functional fixes that can simply be verified through unit tests.

To address these challenges, we introduce \approachName{}, a benchmark specifically designed to evaluate software engineering agents on performance bug fixing tasks in .NET applications. Our benchmark comprises 81 carefully curated and manually verified tasks from popular open-source .NET repositories, each representing a real performance issue that developers fixed in these repos. 


Our evaluation of state-of-the-art coding agents reveals significant challenges in performance optimization tasks. The baseline OpenHands agent achieves only 3\% success rate on \approachName{} with GPT-4.1, substantially lower than its performance on functional bug fixing benchmarks such as SWEBench-Verified (>60\%). 
To demonstrate the potential for improvement, we develop OpenHands-Perf-Agent, an enhanced version that incorporates performance-aware tooling and instructions, achieving $\sim$20\%, much higher compared to baseline. Despite the improvement, we believe significant room for advancement in performance-aware agents still remains.

Our contributions are as follows:
\begin{itemize}
    \item \textbf{\approachName{} Benchmark}: A curated and expert verified collection of 81 real-world performance bug fixing tasks from popular .NET repositories on GitHub. We publicly release the complete benchmark to encourage future research in performance optimization and performance bug fixing in software engineering agents.\footnote{Benchmark available at: \url{https://github.com/glGarg/PerfBench}}
    \item \textbf{Novel Collection/Evaluation Framework}: We share the design of our automated harness that validates performance improvements through agent-generated benchmarks and comparative analysis. Having this kind of harness also allows us to collect commits without needing to have performance test changes within the commit during collection, unlike how functional bugs are typically collected for benchmarks like SWEBench.
    \item \textbf{Empirical Analysis \& Perf Agent}: To demonstrate various limitations of Software Engineering agents today, we build a specialized agent for performance bug-fixing tasks. Through our empirical analysis, we should that this agent has drastically better performance on this benchmark.
\end{itemize}

\section{Background and Related Work}
\label{sec:background}

We discuss work related to performance bug detection and repair, benchmarks for code generation, and software engineering agents.

\subsection{Performance Bugs in Software Systems}

Performance bugs represent a unique class of software defects that impact efficiency without causing functional failures. They tend to be harder to detect ~\cite{attariyan2012x, dean2014perfscope, perfscope, catchmeifyoucan} and fix ~\cite{caramelnistor, song2014oopsla} than functional bugs. Due to being hard to detect and not causing outright failures, these bugs can often go undetected for long periods of time ~\cite{perfscope, catchmeifyoucan}. As a result, better tool support is needed to fix performance bugs. While performance profiling have been developed to identify these issues~\cite{perfscope, attariyan2012x, han-icse2012}, these bugs require significant manual analysis to translate findings into fixes.

\subsection{Automated Program Repair for Performance Issues}

Traditional automated program repair (APR) techniques focus primarily on functional correctness using test suites or logic assertions as specifications. GenProg~\cite{genprog} uses genetic programming to generate fixes that pass test cases, while pattern-based approaches like PAR~\cite{auto_patch_gen} leverage manually created fix templates. Similarly, CapGen~\cite{capgen} generates patches at finer granularity using context-aware prioritization, and VarFix~\cite{varfix} extends GenProg through variational execution~\cite{variexec}. However, these traditional APR approaches face significant challenges when applied to performance bugs. Performance issues require different validation mechanisms than functional correctness, as they must demonstrate measurable improvements in efficiency metrics rather than passing test cases. Additionally, performance bugs often involve subtle inefficiencies that require domain-specific knowledge about algorithms, data structures, and system behavior that is difficult to encode in traditional search-based approaches. Not only that, performance bugs manifest in various forms including algorithmic inefficiencies, memory leaks, excessive allocations, and suboptimal API usage patterns. As a result, for a fix approach to be useful, it would need to be general enough to fix a wide-range of issues. However, many existing approaches target specific kinds of issues such as repeated computations \cite{memoization}, software misconfigurations \cite{misconfigurations}, loop inefficiencies \cite{caramelnistor}. Recent work in model training has explored performance-specific repair. Studies like DeepDev-PERF~\cite{FSEPerf}, RAPGen \cite{garg2025rapgenapproachfixingcode} aim to solve this problem by attempting to fix a wide range of performance issues, using the same model instead of building an approach specialized to a specific category of issues. However, these approaches operate on isolated code snippets rather than full repository contexts that modern agents must handle.

\subsection{Benchmarks for Code Generation and Repair}

The evolution of coding benchmarks has progressed from simple algorithmic tasks to complex real-world scenarios. Early benchmarks like HumanEval~\cite{chen2021evaluating} focused on isolated function generation, achieving high success rates (85-95\%) but limited real-world applicability. The introduction of SWE-bench~\cite{jimenez2024swebench} marked a significant advancement by evaluating agents on real GitHub issues. Current state-of-the-art systems achieve varying performance: Claude models reach >50\% on SWE-bench Verified, while more complex systems like AutoCodeRover achieve 18.83\% on the full benchmark~\cite{zhang2024autocoderover}. However, contamination concerns led to SWE-bench+~\cite{aleithan2024swebenchenhanced}, where performance drops dramatically to 0.55-12\%, revealing the importance of rigorous evaluation.

Recent specialized benchmarks have emerged for specific domains. DevBench~\cite{li2024promptinglargelanguagemodels} evaluates comprehensive software development capabilities, while BigCodeBench~\cite{zhuo2025bigcodebench} focuses on complex programming tasks requiring library usage. RefactorBench ~\cite{related-refactorbench} focuses on hand-crafted refactoring tasks in python repositories. However, none specifically target performance optimization tasks.

\subsection{Language Model Agents for Software Engineering}

Recent advances in coding agents have demonstrated impressive capabilities across software engineering tasks. SWE-agent~\cite{yang2024sweagent} introduced specialized Agent-Computer Interfaces that significantly improved performance on repository-level tasks. OpenHands~\cite{wang2024opendevinopenplatformai} provides a comprehensive platform achieving 29\% success on SWE-bench Full and >60\% on SWE-bench Verified with certain models. Several proprietary systems like Claude Code~\cite{claude_code_2024}, Copilot Agent~\cite{copilot_agent_2024}, Windsurf~\cite{windsurf_2024}, Devin~\cite{cognition2024devin}, etc. have been proposed and demonstrate impressive capabilities in fixing functional bugs. Recent work by Xia et al.~\cite{xia2024agentlessdemystifyingllmbasedsoftware} challenges the necessity of complex agent architectures, showing that simple three-phase approaches can achieve competitive performance (27.3\%) at dramatically lower costs. This suggests that for specialized tasks like performance optimization, targeted approaches may be more effective than general-purpose agent frameworks. However, current agent evaluation focuses almost exclusively on functional bugs when it comes to bug-fixing. We believe that performance bugs require fundamentally different reasoning patterns, including understanding computational complexity, resource utilization, and performance trade-offs. Agents must also master performance-specific tooling and benchmarking methodologies to validate improvements, capabilities that have not been systematically evaluated.

\subsection{Performance Optimization and LLM Applications}

Limited work exists in applying LLMs to performance optimization. Traditional performance optimization relies heavily on profiling tools and expert knowledge. Frameworks like BenchmarkDotNet~\cite{benchmarkdotnet} for .NET provide comprehensive performance measurement capabilities, but require significant expertise to interpret results and guide optimization efforts. Some recent work has explored LLM-based approaches to performance issues. DeepDev-PERF~\cite{FSEPerf} demonstrates a deep learning-based approach for improving software performance, while PerfLens~\cite{garg2021perflens} provides a data-driven performance bug detection and fix platform. However, these approaches operate on isolated methods rather than full repository contexts that modern agents must handle. Research has also developed retrieval-augmented prompt generation approaches for performance bug fixing. Recent work demonstrates that leveraging knowledge-bases of past performance fixes can improve LLM-generated solutions through targeted prompt engineering~\cite{liu2023pre, garg2025rapgenapproachfixingcode}. However, this work operates in controlled settings with pre-identified performance issues rather than the complex repository navigation and issue identification, tasks that real-world agents must perform. Our work bridges these areas by creating an automated framework where agents can generate and validate performance improvements through self-designed benchmarks in realistic repository contexts, addressing the gap between isolated performance optimization and comprehensive software engineering agent evaluation for repo-level improvements.

\section{PerfBench Construction}
\label{sec:benchmark}
Below we explain our data collection process, task design as well as an analysis of examples within \approachName{}. 

\subsection{Data Collection Process}

\begin{figure}[h]
    \centering
    \includegraphics[width=\linewidth]{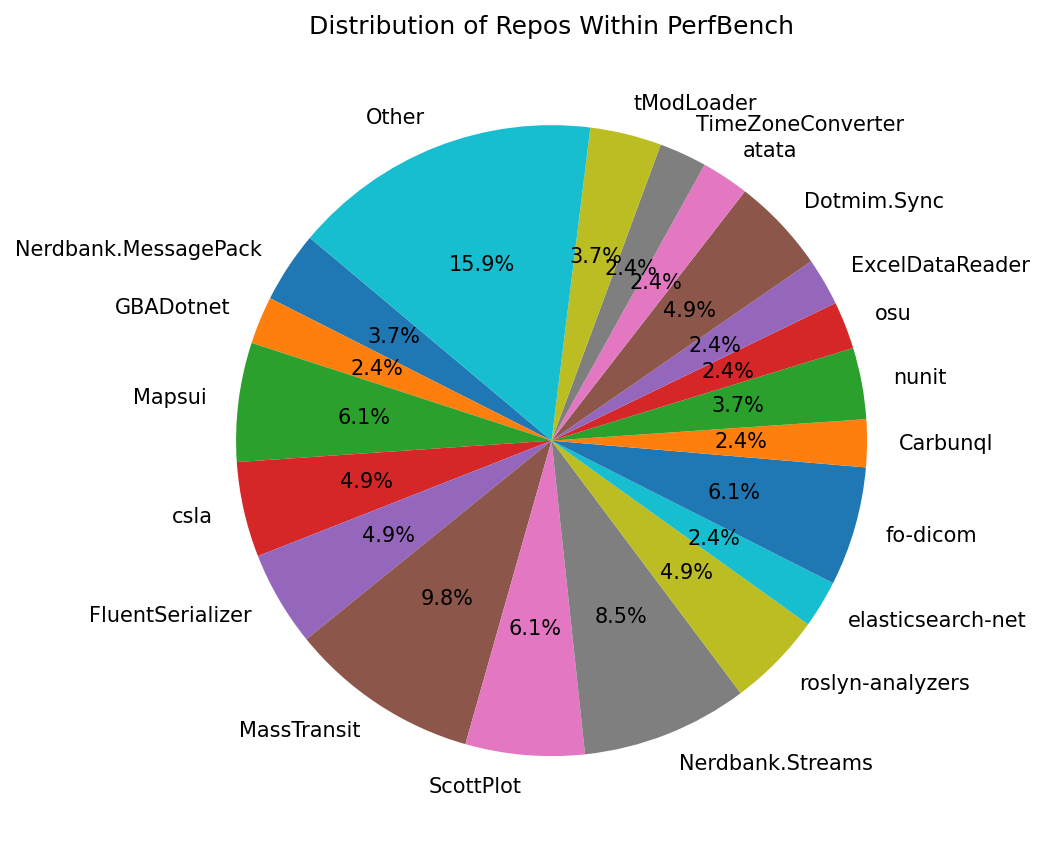}
    \caption{Distribution of C\# repositories in the benchmark.}
    \label{fig:benchmark_repos}
\end{figure}

We collected performance bug fixes through a systematic process. We first identified $\sim$1200 .NET repositories on GitHub with >10 stars, focusing on actively maintained projects with substantial codebases. Within the selected repositories, we crawl the commit history and find commits containing performance-related keywords: "performance", "slow", "optimization", "memory", "CPU", "latency", "throughput", etc. A complete list of all the keywords is provided in Table~\ref{tab:perf_keywords}. We then identified commits that reference an issue and modify at least one .cs file i.e. code changes excluding documentation or configuration changes. For each candidate repo, we create a Docker environment by installing the appropriate version of .NET, available in the .csproj files and verify that both the buggy and fixed versions compile successfully using the \texttt{dotnet build} command. Finally, two experienced .NET developers independently reviewed each issue description and associated fix and issue to confirm that it represents a genuine performance bug and improvement. This process yielded 81 high-quality tasks spanning diverse set of repositories and performance issue types. Figure \ref{fig:benchmark_repos} shows the distribution of repos within our benchmark. We can see that unlike benchmarks like SWE-Bench \cite{jimenez2024swebench}, our benchmark isn't concentrated within a small set of popular python repos. Instead we draw examples from 32 different repos from various domains.

\begin{table*}[h]
\footnotesize
\centering
\caption{Performance-Related Keywords Used for Initial Issue Filtering}
\label{tab:perf_keywords}
\begin{tabular}{p{0.25\textwidth}p{0.7\textwidth}}
\toprule
\textbf{Category} & \textbf{Keywords} \\
\midrule
Core Performance & perf, performance, optimize, optimization, faster, slower, speed up, slow down, latency, throughput, overhead, efficiency, scalable, scalability, bottleneck, lag, load time \\
\midrule
Memory \& CPU & memory, alloc, allocation, dealloc, leak, heap, stack, gc, garbage collection, cpu, utilization, cache, cache miss, oom, out of memory, boxing \\
\midrule
Execution \& Runtime & hot path, critical path \\
\midrule
User-Perceived Issues & hang, freeze, unresponsive, laggy, delay \\
\midrule
Performance Signals & inefficient, excessive, high cpu, high memory, not scalable, timing issue \\
\bottomrule
\end{tabular}
\end{table*}

\definecolor{headerblue}{RGB}{44,90,160}
\definecolor{lightblue}{RGB}{240,248,255}
\definecolor{lightpink}{RGB}{255,245,245}
\definecolor{pinkborder}{RGB}{254,202,202}
\definecolor{darkgray}{RGB}{26,26,26}
\definecolor{errored}{RGB}{255,107,107}

\begin{figure}[htbp]
\centering
\begin{tikzpicture}[
    box/.style={rectangle, draw=black, thick, minimum width=0.9*\columnwidth},
    header/.style={box, fill=headerblue, text=white, minimum height=0.6cm},
    section/.style={box, fill=gray!10, minimum height=0.6cm, text width=0.9*\columnwidth-0.2cm},
    repo/.style={box, fill=lightblue, minimum height=1cm, text width=0.9*\columnwidth-0.2cm},
    issue/.style={box, fill=lightpink, draw=pinkborder, minimum height=3cm, text width=0.9*\columnwidth-0.2cm}
]

\node[header] (header) at (0,0) {\small\textbf{PerfBench Task: codebude\_qrcoder\_\_207\_\_8.0}};

\node[section, below=0.05cm of header] (repo_header) {\footnotesize\textbf{Repository Context}};
\node[repo, below=0.02cm of repo_header] (repo_content) {
    \begin{minipage}{\columnwidth-0.4cm}
    \scriptsize
    \texttt{Repository: codebude/QRCoder} \\
    \texttt{.NET Version: net8.0} \\
    \texttt{Before Commit: eedff45121e33b...} \\
    \texttt{After Commit: 9b135ae6720d9a...}
    \end{minipage}
};

\node[section, below=0.05cm of repo_content] (issue_header) {\footnotesize\textbf{Issue Description}};
\node[issue, below=0.02cm of issue_header] (issue_content) {
    \begin{minipage}{\columnwidth-0.4cm}
    \scriptsize
    \textbf{Expected Behavior} \\
    Returns QR code, which is then printed as graphic object. \\[0.1cm]
    
    \textbf{Current Behavior} \\
    Crash with out-of-memory exception. \\[0.1cm]
    
    \textbf{Steps to Reproduce} \\
    ...\\
    
    \textbf{Environment} \\
    Windows 7. Version used: latest master, as of 15.10.2020... \\[0.1cm]
    
    \textbf{Stack Trace:} \\\\
    \colorbox{darkgray}{\color{errored}\tiny\texttt{
    \begin{minipage}{\columnwidth-0.6cm}
    System.OutOfMemoryException: ... \\
    \phantom{...}v System.Linq.Lookup`2.GetGrouping(TKey key, Boolean create) \\
    \phantom{...}v System.Linq.Lookup`2.Create[TSource](...) \\
    \phantom{...}v QRCoder.QRCodeGenerator.BinaryStringToBitBlockList(...) \\
    \phantom{...}v QRCoder.QRCodeGenerator.GenerateQrCode(...)
    \end{minipage}
    }}
    \end{minipage}
};

\end{tikzpicture}
\caption{Example task from PerfBench showing the metadata collected in our benchmark. The task given to the agent includes repository context and the original GitHub issue describing an OutOfMemoryException.}
\label{fig:task_example}
\end{figure}

\subsection{Extracted Metadata}

Each example within \approachName{} contains a specific set of metadata collected to provide agents with the necessary context for mimicking a real-world development scenario. The metadata collected for each example includes: the repo name, the performance issue description, the commit hashes for the buggy and fixed versions of the code as well as the ground truth patch. Each task is setup to run in an isolated docker container. Figure~\ref{fig:task_example} illustrates a representative task from our benchmark, showing the metadata collected.

\subsection{Task Structure}
The task given to the agent provides access to the complete .NET repository at a specific commit prior to the performance fix from the developer was applied. To ensure a controlled and reproducible environment, we create a Docker image containing the appropriate version of .NET runtime and SDK corresponding to the target project's requirements. The repository is cloned and rolled back to the commit hash before the fix. Note that we also delete the .git folder after checkout to prevent agents from accessing commit history or using git commands to "cheat" by looking at the developer fix. This ensures agents must solve the performance issue independently based solely on the provided repo context.

\subsection{Repository and Commit Statistics}
\label{sec:stats}

To better understand the characteristics of the collected tasks, we analyze the collected GitHub issues, repositories in which the performance issues occurs as well as the gold patch commit. Table~\ref{tab:repo_commit_stats} summarizes key statistics across the 81 tasks in \approachName{}.

\begin{table}[h]
\small
\centering
\caption{Summary statistics of GitHub issues, repositories and gold patch fix in \approachName{}.}
\label{tab:repo_commit_stats}
\begin{tabular}{lccc}
\toprule
\textbf{Metric} & \textbf{Mean} & \textbf{Median} & \textbf{Max} \\
\midrule
Files per Repository & 1{,}227 & 782 & 5{,}595 \\
Words per Problem Statement & 121.6 & 79 & 1{,}006 \\
Modified .cs Files per Fix Commit & 3.4 & 2 & 22 \\
Lines Changed per Fix Commit & 103.7 & 60 & 931 \\
\bottomrule
\end{tabular}
\end{table}

We observe that repositories in \approachName{} are substantially sized, with a mean of nearly >1k files, ensuring that agents must reason through large and realistic codebases. We also look at some statistics from the gold patches. We see that performance bug fix can span up to 22 files and on average >3 files and >100 lines, demonstrating that these issues and fixes are non-trivial.

\subsection{Performance Bug Taxonomy}
\label{sec:taxonomy}
To better understand the types of performance issues represented in \approachName{}, we conducted a systematic analysis of all 81 problem statements. We categorized each issue into a hierarchical taxonomy consisting of high-level categories as well as more granular low-level subcategories. The categorization process involved analyzing the problem descriptions described in the GitHub issues collected and the developer fix and grouping the issues into similar categories.

Our analysis reveals five major categories of performance bugs, with memory management issues being the most prevalent, followed by concurrency problems, algorithmic inefficiencies, I/O performance issues, and build tool performance problems. Table~\ref{tab:perf_taxonomy} presents the complete taxonomy with counts for each category.

The distribution shows that memory-related issues account for >40\% of all performance bugs in our benchmark, highlighting the critical importance of proper memory management and low allocations in .NET applications. Concurrency and algorithmic inefficiency issues each represent approximately 17\% of the benchmark, while I/O and build / test performance issues comprise the remaining cases. In our later analysis, we report the relative performances of agent and model configurations on each category.

\begin{table}[h]
\small
\centering
\caption{Taxonomy of performance bug types in \approachName{}}
\label{tab:perf_taxonomy}
\begin{tabular}{lc}
\toprule
\textbf{Performance Bug Category} & \textbf{Count} \\
\midrule
\textbf{Memory Management Issues} & \textbf{33} \\
\quad Excessive Allocations & 18 \\
\quad Memory Leaks & 15 \\
\midrule
\textbf{Concurrency and Threading Issues} & \textbf{14} \\
\quad Deadlocks and Infinite Loops & 5 \\
\quad Async/Await Issues & 5 \\
\quad Incorrect Parallelism & 4 \\
\midrule
\textbf{Inefficient Algorithms and Data Structures} & \textbf{14} \\
\quad Algorithmic Inefficiency & 9 \\
\quad .NET Collection-related Performance Issues & 4 \\
\quad Other Data Structure Misuse & 1 \\
\midrule
\textbf{I/O and Serialization Performance} & \textbf{12} \\
\quad Serialization Inefficiency & 10 \\
\quad Network I/O Issues & 2 \\
\midrule
\textbf{Build and Analysis Tools Performance} & \textbf{8} \\
\quad Test Performance & 4 \\
\quad Build Performance & 4 \\
\midrule
\textbf{Total} & \textbf{81} \\
\bottomrule
\end{tabular}
\end{table}

\section{Experimental Setup}

\subsection{Evaluation Harness Design}
The \approachName{} evaluation harness automates the entire testing process. It extracts benchmark tests generated by the agent itself as well as the code changes containing performance improvements suggested by the agent. Benchmarks for .NET apps are typically written using the BenchmarkDotNet \cite{benchmarkdotnet} framework. Our harness expects the agents to have written benchmark tests using the standard framework instead of rolling its own benchmarking code. This is a standard practice followed by performance engineers in .NET. We then execute the agent-written BenchmarkDotNet tests before and after the changes suggested by the agent and developer along with executing unit tests present within the repo. We report the success rate as well as several other metrics described below.

\begin{figure*}[htbp]
\small
\centering
\colorbox{black}{%
\begin{minipage}{0.99\textwidth}
\color{white}
\fontfamily{cmtt}\selectfont
\footnotesize

\vspace{0.3cm}
BenchmarkDotNet v0.13.12, OS=Windows 11.0.22621.2428 (22H2/2022Update/SunValley2)\\
Intel Core i9-12900K CPU 3.20GHz (Alder Lake), 1 CPU, 24 logical and 16 physical cores\\
\phantom{}\quad[Host]\phantom{}\phantom{}\phantom{}\phantom{}\phantom{}\phantom{}\phantom{}: .NET 8.0.0 (8.0.23.53103), X64 RyuJIT AVX2\\
\phantom{}\quad.NET 8.0\phantom{}\phantom{}\phantom{}\phantom{}\phantom{}: .NET 8.0.0 (8.0.23.53103), X64 RyuJIT AVX2\\
\phantom{}\quad.NET Framework : .NET Framework 4.8.1 (4.8.9181.0), X64 RyuJIT\\

\vspace{0.2cm}

\begin{tabular}{|l|r|r|r|r|r|r|r|r|r|}
\hline
\textbf{Method} & \textbf{Mean} & \textbf{StdDev} & \textbf{StdErr} & \textbf{Median} & \textbf{Q1} & \textbf{Q3} & \textbf{Iterations} & \textbf{Gen 0} & \textbf{Allocated} \\
\hline
JsonSerialization & 1.247 ms & 0.0423 ms & 0.0089 ms & 1.239 ms & 1.218 ms & 1.273 ms & 50 & 0.1953 & 1,024 B \\
\hline
XmlSerialization & 2.891 ms & 0.0891 ms & 0.0213 ms & 2.876 ms & 2.834 ms & 2.945 ms & 35 & 0.4883 & 2,512 B \\
\hline
\end{tabular}

\vspace{0.3cm}
\end{minipage}%
}
\caption{A sample BenchmarkDotNet output table showing execution time statistics, memory allocation and Garbage Collection (GC) metrics across different tests within the test suite. Our evaluation harness uses this to determine whether the code changes improve performance.}
\label{fig:benchmarkdotnet-results}
\end{figure*}

\subsection{Metrics}
For each task, the harness computes the following metrics:
\begin{itemize}
    \item \textbf{Success Rate (\%)}: Percentage of tasks where the agent produces a fix that improves performance on at least one benchmark without causing regressions to other benchmarks, as well as passing all existing unit tests in the repo.
    \item \textbf{Performance Improvement (\%, kbs, ms, etc.)}: For successful fixes, we use the summary statistics table output at the end of a BenchmarkDotNet test (Figure \ref{fig:benchmarkdotnet-results}) to measure the Execution time reduction, Memory usage reduction (kbs), depending on the type of performance improvement.
    \item \textbf{Token Usage (\# tokens)}: In addition to correctness metrics, the harness also shows the number of input and output tokens taken by the agent to solve the task. Since tokens are proportional to cost of using the LLM, we want this to be as low as possible.
    \item \textbf{Steps Taken (\# steps)}: We also report the number of steps taken by the agent to solve the task. Steps translate to latency, so we want this to be low for high responsiveness.
    \item \textbf{Dollar Cost (\$)}: Finally, we report the cost of running the agent per instance based on cost-per-million numbers provided by the LLM provider.
\end{itemize}

\subsection{Agent Configurations}

We evaluate the open-source Software Engineering agent OpenHands \cite{wang2024opendevinopenplatformai} against the benchmark in two different configurations:

\begin{itemize}
    \item \textbf{OpenHands (Baseline)}: As our baseline, we use the OpenHands agent with default prompting for bug fixing tasks with the only change being to update the prompt to target C\# instead of python. This baseline represents the current state-of-the-art general-purpose coding agents, which we find are targeted mainly towards functional bugs.
    \item \textbf{OpenHands-Perf-Agent}: We create a fork of OpenHands agent with changes specific to performance. The changes include: performance-aware instructions and planning such as explicit benchmark generation instructions for the LLM, and benchmark output processing. We discuss this in more detail in the next section.
\end{itemize}

We run each agent with two state-of-the-art language models to evaluate performance in different families of closed-source models. For cost purposes, we limit our evaluations to the following two models: GPT-4.1 and Claude Sonnet 4.

Each agent is allowed a maximum of 100 steps per task to mimic realistic usage without being too expensive and time-consuming, while still providing ample opportunity for the agent to iterate over the problem. Tasks are executed in isolated Docker containers to ensure reproducibility and prevent interference between runs.

\subsection{A Perf Agent}
We modify the OpenHands agent to create OpenHands-Perf-Agent with specific enhancements for performance optimization tasks. The two key modifications include:

\subsubsection{Performance-Aware \& Benchmarking Instructions}
We modify the instructions to explicitly guide the agent through the desired sequence of high-level steps for performance optimization workflows. We also provide the agent specific instructions for creating BenchmarkDotNet tests with appropriate diagnostics for measuring memory usage, knowing how important memory is for .NET applications. Figure~\ref{fig:prompt_diff} shows the key differences between the baseline OpenHands prompt and our performance-optimized version.

\subsubsection{Output Processing}
We also add tooling with custom parsing of benchmark results to extract relevant performance metrics to avoid token overflow. This is a critical enhancement in OpenHands-Perf-Agent as the raw output from performance benchmarks can be extremely verbose (often exceeding 10k tokens for a single execution), which quickly overwhelms LLMs due to their context limitations. Our solution implements the following output parsing:

\begin{figure}[h]
\centering
\begin{minipage}{\columnwidth}
\scriptsize
\ttfamily
\setlength{\fboxsep}{2pt}
\setlength{\parskip}{1pt}
\noindent I've uploaded a C\# code repository in the\\
directory \{workspace\_dir\}.\\
\colorbox{red!20}{Consider the following issue description:}\\
\colorbox{green!20}{Consider the following perf issue description:}\\
\noindent \{issue\_description\_text\}\\
\noindent Can you help me implement the necessary\\
changes to the repository so that the requirements\\
specified in the <issue\_description> are met?\\
\noindent Follow these steps to resolve the issue:\\
1. As a first step, it might be a good idea to explore\\
\phantom{1. }the repo to familiarize yourself with its structure.\\
\colorbox{red!20}{2. Create a script to reproduce the error and execute}\\
\colorbox{red!20}{\phantom{2. }it with `dotnet run` or `dotnet build` using the}\\
\colorbox{red!20}{\phantom{2. }BashTool, to confirm the error}\\
\colorbox{green!20}{2. Create Benchmark Tests: Use BenchmarkDotNet +}\\
\colorbox{green!20}{\phantom{2. }MemoryDiagnoser. Create .csproj and run the tests}\\
\colorbox{green!20}{\phantom{2. }using `dotnet run` command with the BashTool.}\\
\colorbox{red!20}{3. Edit the source code of the repo to resolve the issue}\\
\colorbox{green!20}{3. Edit the source code to optimize the code in the repo.}\\
\colorbox{red!20}{4. Re-run your reproduce script and confirm that the}\\
\colorbox{red!20}{\phantom{4. }error is fixed!}\\
\colorbox{green!20}{4. Re-run your benchmark and confirm that the}\\
\colorbox{green!20}{\phantom{4. }performance has improved!}\\
\colorbox{red!20}{5. Think about edgecases and make sure your fix}\\
\colorbox{red!20}{\phantom{5. }handles them as well}\\
\colorbox{green!20}{5. Run any unit tests in the repo to ensure correctness}\\
\colorbox{green!20}{\phantom{5. }of your changes.}\\
\colorbox{green!20}{6. Finally explain the fix you implemented and output}\\
\colorbox{green!20}{\phantom{6. }a markdown with a description of the changes and}\\
\colorbox{green!20}{\phantom{6. }include tables with benchmark results.}\\
\end{minipage}
\caption{Prompt template differences between baseline OpenHands (light red) and OpenHands-Perf-Agent (light green). The performance-aware version replaces error reproduction with benchmark creation and emphasizes the high-level sequence of steps we expect the agent to follow to fix a performance bug.}
\label{fig:prompt_diff}
\end{figure}

\begin{itemize}
    \item \textbf{Success Case}: When benchmarks execute successfully, i.e. we see a table at the end of the output, we extract only the summary table (shown in Figure \ref{fig:benchmarkdotnet-results}) containing key perf metrics and discard verbose diagnostics preceding it.
    \item \textbf{Error Case}: When execution fails, we preserve the complete output to enable agent to diagnose and rectify issues.
\end{itemize}

We measured that this approach reduces token usage associated with benchmark outputs by >90\%, while preserving essential performance information.

\section{Benchmark Results}
\label{sec:results}

\subsection{Overall Performance}

Table~\ref{tab:main_results} presents the main experimental results of our two agent configurations on \approachName{} with the LLM models we use.

\begin{table*}[h]
\footnotesize
\centering
\caption{Overall performances of different agents and model configurations on \approachName{}}
\label{tab:main_results}
\begin{tabular}{llcccc}
\toprule
\textbf{Agent} & \textbf{Model} & \textbf{Success Rate (\%)} & \textbf{\# of Avg Steps} & \textbf{\# of Avg Tokens} & \textbf{Avg Cost (\$)} \\
\midrule
\multirow{3}{*}{OpenHands (Baseline)} 
& GPT-4.1 & 1.2\% (1/81) & 47.2 & 1.3M & 15.42 \\
& Claude Sonnet 4 & 3.7\% (3/81) & 62.3 & 2.6M & 7.92 \\
\midrule
\multirow{3}{*}{OpenHands-Perf-Agent} 
& GPT-4.1 & 14.8\% (12/81) & 84.3 & 1.9M & 23.08 \\
& Claude Sonnet 4 & 19.7\% (16/81) & 49.0 & 1.7M & 5.03 \\
\bottomrule
\end{tabular}
\end{table*}

\begin{table*}[h]
\footnotesize
\centering
\caption{Resolution rates by high-level performance bug categories in Table \ref{tab:perf_taxonomy} of our Baseline OpenHands and OpenHands-Perf-Agent configurations with Claude Sonnet 4}
\label{tab:category_results}
\begin{tabular}{lccc}
\toprule
\textbf{Category} & \textbf{Total} & \textbf{Baseline} & \textbf{Perf-Agent} \\
\midrule
Memory Management & 33 & 6.0\% (2/33) & 18.2\% (6/33)\\
Concurrency \& Threading & 14 & 0\% (0/14) & 14.3\% (2/14)\\
Inefficient Algorithms & 14 & 7.1\% (1/14) & 21.4\% (3/14)\\
I/O \& Serialization & 12 & 0\% (0/12) & 33.3\% (4/12)\\
Build \& Analysis Tools & 9 & 0\% (0/8) & 12.5\% (1/8)\\
\midrule
\textbf{Total} & \textbf{81} & \textbf{3.7\% (3/81)} & \textbf{19.7\% (16/81)}\\
\bottomrule
\end{tabular}
\end{table*}

The results reveal significant challenges in performance bug fixing for current LM agents. The baseline OpenHands agent achieves <4\% success rate across both models, substantially lower than typical performance on functional bug fixing benchmarks such as SWE-bench Verified (>60\%~\cite{wang2024opendevinopenplatformai}). This dramatic performance gap highlights the fundamental differences between functional and performance bug fixing tasks.

Our performance-aware agent demonstrates an improvement, achieving 15-20\% success rate, with up to 5x improvement over the baseline depending on the model.

\begin{figure}[htbp]
\centering
\resizebox{\columnwidth}{!}{
\begin{tikzpicture}
\begin{axis}[
    width=0.9\columnwidth,
    height=2.2cm,
    boxplot/draw direction=x,
    xlabel={CPU Improvement (\%)},
    ytick={1},
    yticklabels={CPU},
    yticklabel style={text width=2.5cm, align=right},
    ymin=0.5, ymax=1.5,
    xmin=0, xmax=100,
    grid=major,
    grid style={gray!30},
    boxplot={
        draw position=1,
        box extend=0.3,
        whisker extend=0.15,
    },
    boxplot/every box/.style={fill=blue!50, fill opacity=0.6, draw=blue!80},
    boxplot/every whisker/.style={draw=blue!80},
    boxplot/every median/.style={draw=blue!80, line width=1.5pt},
]
\addplot+[
    boxplot prepared={
        median=2.459,
        upper quartile=42.902,
        lower quartile=1.106,
        upper whisker=90.383,
        lower whisker=0.613,
    },
] coordinates {};
\end{axis}
\end{tikzpicture}
}
\vspace{0.05cm}

\resizebox{\columnwidth}{!}{
\begin{tikzpicture}
\begin{axis}[
    width=0.9\columnwidth,
    height=2.2cm,
    boxplot/draw direction=x,
    xlabel={CPU Improvement ($\mu$s)},
    ytick={1},
    yticklabels={CPU (abs)},
    yticklabel style={text width=2.5cm, align=right},
    ymin=0.5, ymax=1.5,
    xmode=log,
    log basis x=10,
    xmin=0.1, xmax=1000,
    xtick={0.1, 1, 10, 100, 1000},
    xticklabels={$10^{-1}$, $10^0$, $10^1$, $10^2$, $10^3$},
    grid=major,
    grid style={gray!30},
    boxplot={
        draw position=1,
        box extend=0.3,
        whisker extend=0.15,
    },
    boxplot/every box/.style={fill=blue!30, fill opacity=0.6, draw=blue!60},
    boxplot/every whisker/.style={draw=blue!60},
    boxplot/every median/.style={draw=blue!60, line width=1.5pt},
]
\addplot+[
    boxplot prepared={
        median=9.300,
        upper quartile=166.600,
        lower quartile=8.306,
        upper whisker=385.97,
        lower whisker=0.42,
    },
] coordinates {};


\end{axis}
\end{tikzpicture}
}
\vspace{0.05cm}

\resizebox{\columnwidth}{!}{
\begin{tikzpicture}
\begin{axis}[
    width=0.9\columnwidth,
    height=2.2cm,
    boxplot/draw direction=x,
    xlabel={Memory Improvement (\%)},
    ytick={1},
    yticklabels={Memory},
    yticklabel style={text width=2.5cm, align=right},
    ymin=0.5, ymax=1.5,
    xmin=0, xmax=100,
    grid=major,
    grid style={gray!30},
    boxplot={
        draw position=1,
        box extend=0.3,
        whisker extend=0.15,
    },
    boxplot/every box/.style={fill=red!50, fill opacity=0.6, draw=red!80},
    boxplot/every whisker/.style={draw=red!80},
    boxplot/every median/.style={draw=red!80, line width=1.5pt},
]
\addplot+[
    boxplot prepared={
        median=70.558,
        upper quartile=81.479,
        lower quartile=23.752,
        upper whisker=87.796,
        lower whisker=0.024,
    },
] coordinates {};
\end{axis}
\end{tikzpicture}
}
\vspace{0.05cm}

\resizebox{\columnwidth}{!}{
\begin{tikzpicture}
\begin{axis}[
    width=0.9\columnwidth,
    height=2.2cm,
    boxplot/draw direction=x,
    xlabel={Memory Improvement (bytes)},
    ytick={1},
    yticklabels={Memory (abs)},
    yticklabel style={text width=2.5cm, align=right},
    ymin=0.5, ymax=1.5,
    xmode=log,
    log basis x=2,
    xmin=1, xmax=1048576,
    xtick={1, 32, 1024, 32768, 1048576},
    xticklabels={$2^0$, $2^5$, $2^{10}$, $2^{15}$, $2^{20}$},
    grid=major,
    grid style={gray!30},
    boxplot={
        draw position=1,
        box extend=0.3,
        whisker extend=0.15,
    },
    boxplot/every box/.style={fill=red!30, fill opacity=0.6, draw=red!60},
    boxplot/every whisker/.style={draw=red!60},
    boxplot/every median/.style={draw=red!60, line width=1.5pt},
]
\addplot+[
    boxplot prepared={
        median=7244.800,
        upper quartile=236193.280,
        lower quartile=2119.680,
        upper whisker=348221.44,
        lower whisker=1,
    },
] coordinates {};
\end{axis}
\end{tikzpicture}
}
\caption{Performance improvements achieved by OpenHands-Perf-Agent. The horizontal boxplots show the distribution of CPU and Memory improvements in both relative (\%) and absolute units ($\mu$s for CPU usage, bytes for allocations) across benchmark tests.}
\label{fig:performance_improvements}
\end{figure}
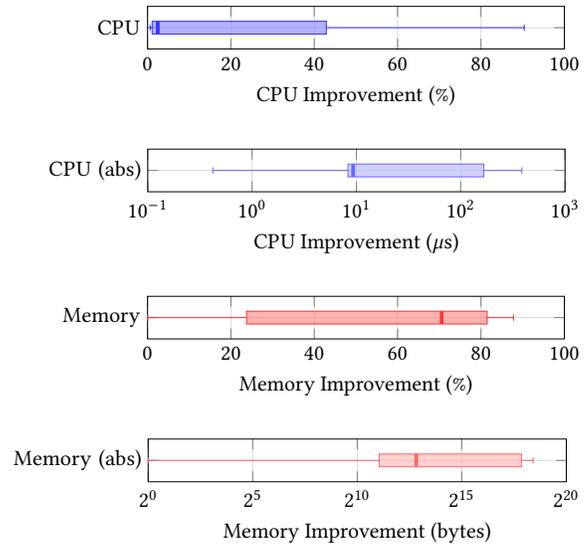

\subsection{Performance Metrics Analysis}

To provide deeper insights into the nature of performance improvements achieved by our agents, we analyze the change in CPU and memory metrics for successful fixes. Figure~\ref{fig:performance_improvements} presents the distribution of absolute and relative improvements in memory allocation and CPU utilization compared to the original code.

Memory allocation reductions (Figure~\ref{fig:performance_improvements} demonstrate the more prominent improvements, with all successful fixes typically having more than 20\% improvement to allocations, with several having improvements on the order of KBs and MBs. Unlike Memory, CPU utilization improvements (Figure~\ref{fig:performance_improvements} are typically in the range from 0 to 40\%, but are on the order of microseconds or milliseconds. The highest CPU improvements correspond to fixes addressing algorithmic inefficiencies.

While improvements on the lower end of the plot may seem like micro-optimizations, we would like to point out that these numbers are taken from individual benchmark tests that represent an isolated execution of a specific code path in the application and don't directly reflect usage of the application by an end-user. Based on whether the fixed code is executed on the application's hot-path, the customer may see significant improvements to performance.

\subsection{Performance by Bug Category}

Table~\ref{tab:category_results} breaks down the performances of OpenHands-Perf-Agent and baseline OpenHands (with Claude Sonnet 4) across the different performance bug categories identified in the high-level categories we found in Section \ref{sec:taxonomy}.

The results show that I/O \& Serialization bugs are most successfully addressed ($\sim$33\% success rate), followed by algorithmic issues ($\sim$21\%). The fact that algorithmic issues have a high pass rate aligns with our understanding that algorithmic improvements often have clear patterns that LLMs can recognize from training data.

Notably, the baseline agent fails completely on most of the categories, achieving only a handful of successful runs in memory management and algorithmic issues. This suggests that without explicit instructions, agents struggle to identify and address complex performance patterns beyond the most obvious inefficiencies. Even the OpenHands-Perf-Agent configuration struggles with other categories such as Build \& Analysis Tools, Concurrency, etc., showing that there is room for much improvement.


\section{Limitations}
\label{sec:limitations}

While \approachName{} provides valuable insights into agent capabilities for perf optimization, our work has various limitations that must be considered.

\textbf{Performance vs. Correctness Trade-off}: Our evaluation harness primarily focuses on measurable performance improvements rather than assessing whether agents implement the correct fix for the underlying performance issue discussed in the GitHub issue. An agent may achieve performance gains through alternative optimizations that differ from the developer's intended solution, yet still receive full credit in our evaluation. While such improvements may be practically valuable, they may not demonstrate true understanding of the root cause identified by the original developer.

\textbf{Dependency on Benchmarking}: Our evaluation relies entirely on agent-generated benchmarks, creating a potential weakness where poorly designed benchmarks may fail to capture the true performance characteristics of the issue. Agents that generate inadequate benchmarks may receive false negatives even with correct fixes. However, we would like to note that writing benchmarks in .NET is a relatively easy task for today's LLMs and easier still compared to writing unit tests which require deeper understanding of the code and coming up with test cases that exercise crucial code paths. On the other hand benchmarks are simply measurement tools that don't require such deep analysis.

\textbf{Language Constraints}: \approachName{} focuses exclusively on .NET applications and C\# code, limiting generalizability of the benchmark and our learnings from building a performance-aware agent to other programming languages. We leave the exploration of other languages and frameworks to future work.

\textbf{Focus on CPU and Memory}: Our evaluation primarily considers improvements in execution time and memory allocation without capturing other important performance dimensions such as network usage, disk I/O usage, or end user-perceived latency.

Despite these limitations, we believe that \approachName{} provides a valuable foundation for evaluating and improving agent capabilities in performance optimization tasks. Future work can address these constraints through expanded language support, more sophisticated evaluation methodologies, and broader metric coverage.

\section{Conclusion}
\label{sec:conclusion}

In this work, we introduced \approachName{}, the first benchmark for evaluating language model agents on real-world performance bug fixing tasks in .NET. Comprising 81 carefully curated tasks from diverse .NET repositories, \approachName{} features a novel evaluation framework that enables agents to generate their own performance benchmarks and validates fixes through comparative metric analysis. Our comprehensive evaluation reveals that current state-of-the-art agents struggle significantly with performance optimization, achieving <4\% success rates compared to the >60\% these same agents typically achieve on functional bug fixing benchmarks. Through OpenHands-Perf-Agent, which incorporates performance-aware instructions and specialized tooling, we demonstrated that targeted approaches can yield substantial improvements, achieving up to 20\% success rates, which is a 5x improvement over baseline agents. However, significant gaps remain between agent and human developer capabilities, particularly for complex categories like concurrency \& threading, and build performance issues.

\approachName{} establishes a challenging benchmark for advancing agents beyond just functional bug fixing into the domain of performance optimization. As software systems increasingly prioritize efficiency and resource optimization in cloud-native and cost-sensitive environments, developing agents capable of performance-aware reasoning becomes essential. Our work reveals that current software engineering agents have not kept pace with this growing need. We hope \approachName{} sparks further research in this important but understudied area.

\bibliographystyle{ACM-Reference-Format}
\bibliography{iclr2024_conference}


\begin{thebibliography}{33}


\ifx \showCODEN    \undefined \def \showCODEN     #1{\unskip}     \fi
\ifx \showDOI      \undefined \def \showDOI       #1{#1}\fi
\ifx \showISBNx    \undefined \def \showISBNx     #1{\unskip}     \fi
\ifx \showISBNxiii \undefined \def \showISBNxiii  #1{\unskip}     \fi
\ifx \showISSN     \undefined \def \showISSN      #1{\unskip}     \fi
\ifx \showLCCN     \undefined \def \showLCCN      #1{\unskip}     \fi
\ifx \shownote     \undefined \def \shownote      #1{#1}          \fi
\ifx \showarticletitle \undefined \def \showarticletitle #1{#1}   \fi
\ifx \showURL      \undefined \def \showURL       {\relax}        \fi
\providecommand\bibfield[2]{#2}
\providecommand\bibinfo[2]{#2}
\providecommand\natexlab[1]{#1}
\providecommand\showeprint[2][]{arXiv:#2}

\bibitem[Aleithan et~al\mbox{.}(2024)]%
        {aleithan2024swebenchenhanced}
\bibfield{author}{\bibinfo{person}{Reem Aleithan}, \bibinfo{person}{Haoran Xue}, \bibinfo{person}{Mohammad~Mahdi Mohajer}, \bibinfo{person}{Elijah Nnorom}, \bibinfo{person}{Gias Uddin}, {and} \bibinfo{person}{Song Wang}.} \bibinfo{year}{2024}\natexlab{}.
\newblock \bibinfo{title}{SWE-Bench+: Enhanced Coding Benchmark for LLMs}.
\newblock
\newblock
\showeprint[arxiv]{2410.06992}~[cs.SE]
\urldef\tempurl%
\url{https://arxiv.org/abs/2410.06992}
\showURL{%
\tempurl}


\bibitem[{Anthropic}(2024)]%
        {claude_code_2024}
\bibfield{author}{\bibinfo{person}{{Anthropic}}.} \bibinfo{year}{2024}\natexlab{}.
\newblock \bibinfo{title}{{Claude for Coding}}.
\newblock \bibinfo{howpublished}{\url{https://www.anthropic.com/claude-code}}.
\newblock
\newblock
\shownote{Accessed: 2025-07-14}.


\bibitem[Attariyan et~al\mbox{.}(2012)]%
        {attariyan2012x}
\bibfield{author}{\bibinfo{person}{Mona Attariyan}, \bibinfo{person}{Michael Chow}, {and} \bibinfo{person}{Jason Flinn}.} \bibinfo{year}{2012}\natexlab{}.
\newblock \showarticletitle{X-ray: Automating $\{$Root-Cause$\}$ Diagnosis of Performance Anomalies in Production Software}. In \bibinfo{booktitle}{\emph{10th USENIX Symposium on Operating Systems Design and Implementation (OSDI 12)}}. \bibinfo{pages}{307--320}.
\newblock


\bibitem[Austin and Flanagan(2012)]%
        {variexec}
\bibfield{author}{\bibinfo{person}{Thomas~H. Austin} {and} \bibinfo{person}{Cormac Flanagan}.} \bibinfo{year}{2012}\natexlab{}.
\newblock \showarticletitle{Multiple Facets for Dynamic Information Flow}.
\newblock \bibinfo{journal}{\emph{SIGPLAN Not.}} \bibinfo{volume}{47}, \bibinfo{number}{1} (\bibinfo{date}{jan} \bibinfo{year}{2012}), \bibinfo{pages}{165–178}.
\newblock
\showISSN{0362-1340}
\urldef\tempurl%
\url{https://doi.org/10.1145/2103621.2103677}
\showDOI{\tempurl}


\bibitem[Chen et~al\mbox{.}(2021)]%
        {chen2021evaluating}
\bibfield{author}{\bibinfo{person}{Mark Chen}, \bibinfo{person}{Jerry Tworek}, \bibinfo{person}{Heewoo Jun}, \bibinfo{person}{Qiming Yuan}, \bibinfo{person}{Henrique~Ponde de Oliveira~Pinto}, \bibinfo{person}{Jared Kaplan}, \bibinfo{person}{Harri Edwards}, \bibinfo{person}{Yuri Burda}, \bibinfo{person}{Nicholas Joseph}, \bibinfo{person}{Greg Brockman}, \bibinfo{person}{Alex Ray}, \bibinfo{person}{Raul Puri}, \bibinfo{person}{Gretchen Krueger}, \bibinfo{person}{Michael Petrov}, \bibinfo{person}{Heidy Khlaaf}, \bibinfo{person}{Girish Sastry}, \bibinfo{person}{Pamela Mishkin}, \bibinfo{person}{Brooke Chan}, \bibinfo{person}{Scott Gray}, \bibinfo{person}{Nick Ryder}, \bibinfo{person}{Mikhail Pavlov}, \bibinfo{person}{Alethea Power}, \bibinfo{person}{Lukasz Kaiser}, \bibinfo{person}{Mohammad Bavarian}, \bibinfo{person}{Clemens Winter}, \bibinfo{person}{Philippe Tillet}, \bibinfo{person}{Felipe~Petroski Such}, \bibinfo{person}{Dave Cummings}, \bibinfo{person}{Matthias Plappert}, \bibinfo{person}{Fotios Chantzis},
  \bibinfo{person}{Elizabeth Barnes}, \bibinfo{person}{Ariel Herbert-Voss}, \bibinfo{person}{William~Hebgen Guss}, \bibinfo{person}{Alex Nichol}, \bibinfo{person}{Alex Paino}, \bibinfo{person}{Nikolas Tezak}, \bibinfo{person}{Jie Tang}, \bibinfo{person}{Igor Babuschkin}, \bibinfo{person}{Suchir Balaji}, \bibinfo{person}{Shantanu Jain}, \bibinfo{person}{William Saunders}, \bibinfo{person}{Christopher Hesse}, \bibinfo{person}{Andrew~N. Carr}, \bibinfo{person}{Jan Leike}, \bibinfo{person}{Josh Achiam}, \bibinfo{person}{Vedant Misra}, \bibinfo{person}{Evan Morikawa}, \bibinfo{person}{Alec Radford}, \bibinfo{person}{Matthew Knight}, \bibinfo{person}{Miles Brundage}, \bibinfo{person}{Mira Murati}, \bibinfo{person}{Katie Mayer}, \bibinfo{person}{Peter Welinder}, \bibinfo{person}{Bob McGrew}, \bibinfo{person}{Dario Amodei}, \bibinfo{person}{Sam McCandlish}, \bibinfo{person}{Ilya Sutskever}, {and} \bibinfo{person}{Wojciech Zaremba}.} \bibinfo{year}{2021}\natexlab{}.
\newblock \bibinfo{title}{Evaluating Large Language Models Trained on Code}.
\newblock
\newblock
\showeprint[arxiv]{2107.03374}~[cs.LG]
\urldef\tempurl%
\url{https://arxiv.org/abs/2107.03374}
\showURL{%
\tempurl}


\bibitem[Cognition.ai(2024)]%
        {cognition2024devin}
\bibfield{author}{\bibinfo{person}{Cognition.ai}.} \bibinfo{year}{2024}\natexlab{}.
\newblock \bibinfo{title}{Introducing DEVIN}.
\newblock
\newblock
\urldef\tempurl%
\url{https://www.cognition.ai/blog/introducing-devin}
\showURL{%
\tempurl}
\newblock
\shownote{Accessed: 2024-08-08}.


\bibitem[Dean et~al\mbox{.}(2014a)]%
        {dean2014perfscope}
\bibfield{author}{\bibinfo{person}{Daniel~J. Dean}, \bibinfo{person}{Hiep Nguyen}, \bibinfo{person}{Xiaohui Gu}, \bibinfo{person}{Hui Zhang}, \bibinfo{person}{Junghwan Rhee}, \bibinfo{person}{Nipun Arora}, {and} \bibinfo{person}{Geoff Jiang}.} \bibinfo{year}{2014}\natexlab{a}.
\newblock \showarticletitle{PerfScope: Practical Online Server Performance Bug Inference in Production Cloud Computing Infrastructures}. In \bibinfo{booktitle}{\emph{Proceedings of the ACM Symposium on Cloud Computing}} (Seattle, WA, USA) \emph{(\bibinfo{series}{SOCC '14})}. \bibinfo{publisher}{Association for Computing Machinery}, \bibinfo{address}{New York, NY, USA}, \bibinfo{pages}{1–13}.
\newblock
\showISBNx{9781450332521}
\urldef\tempurl%
\url{https://doi.org/10.1145/2670979.2670987}
\showDOI{\tempurl}


\bibitem[Dean et~al\mbox{.}(2014b)]%
        {perfscope}
\bibfield{author}{\bibinfo{person}{Daniel~J. Dean}, \bibinfo{person}{Hiep Nguyen}, \bibinfo{person}{Xiaohui Gu}, \bibinfo{person}{Hui Zhang}, \bibinfo{person}{Junghwan Rhee}, \bibinfo{person}{Nipun Arora}, {and} \bibinfo{person}{Geoff Jiang}.} \bibinfo{year}{2014}\natexlab{b}.
\newblock \showarticletitle{PerfScope: Practical Online Server Performance Bug Inference in Production Cloud Computing Infrastructures}. In \bibinfo{booktitle}{\emph{Proceedings of the ACM Symposium on Cloud Computing}} (Seattle, WA, USA) \emph{(\bibinfo{series}{SOCC '14})}. \bibinfo{publisher}{Association for Computing Machinery}, \bibinfo{address}{New York, NY, USA}, \bibinfo{pages}{1–13}.
\newblock
\showISBNx{9781450332521}
\urldef\tempurl%
\url{https://doi.org/10.1145/2670979.2670987}
\showDOI{\tempurl}


\bibitem[Della~Toffola et~al\mbox{.}(2015)]%
        {memoization}
\bibfield{author}{\bibinfo{person}{Luca Della~Toffola}, \bibinfo{person}{Michael Pradel}, {and} \bibinfo{person}{Thomas~R. Gross}.} \bibinfo{year}{2015}\natexlab{}.
\newblock \showarticletitle{Performance Problems You Can Fix: A Dynamic Analysis of Memoization Opportunities}.
\newblock \bibinfo{journal}{\emph{SIGPLAN Not.}} \bibinfo{volume}{50}, \bibinfo{number}{10} (\bibinfo{date}{oct} \bibinfo{year}{2015}), \bibinfo{pages}{607–622}.
\newblock
\showISSN{0362-1340}
\urldef\tempurl%
\url{https://doi.org/10.1145/2858965.2814290}
\showDOI{\tempurl}


\bibitem[Garg et~al\mbox{.}(2022)]%
        {FSEPerf}
\bibfield{author}{\bibinfo{person}{Spandan Garg}, \bibinfo{person}{Roshanak~Zilouchian Moghaddam}, \bibinfo{person}{Colin~B. Clement}, \bibinfo{person}{Neel Sundaresan}, {and} \bibinfo{person}{Chen Wu}.} \bibinfo{year}{2022}\natexlab{}.
\newblock \showarticletitle{DeepDev-PERF: A Deep Learning-Based Approach for Improving Software Performance} \emph{(\bibinfo{series}{ESEC/FSE 2022})}. \bibinfo{publisher}{Association for Computing Machinery}, \bibinfo{address}{New York, NY, USA}, \bibinfo{pages}{948–958}.
\newblock
\showISBNx{9781450394130}
\urldef\tempurl%
\url{https://doi.org/10.1145/3540250.3549096}
\showDOI{\tempurl}


\bibitem[Garg et~al\mbox{.}(2025)]%
        {garg2025rapgenapproachfixingcode}
\bibfield{author}{\bibinfo{person}{Spandan Garg}, \bibinfo{person}{Roshanak~Zilouchian Moghaddam}, {and} \bibinfo{person}{Neel Sundaresan}.} \bibinfo{year}{2025}\natexlab{}.
\newblock \bibinfo{title}{RAPGen: An Approach for Fixing Code Inefficiencies in Zero-Shot}.
\newblock
\newblock
\showeprint[arxiv]{2306.17077}~[cs.SE]
\urldef\tempurl%
\url{https://arxiv.org/abs/2306.17077}
\showURL{%
\tempurl}


\bibitem[Garg et~al\mbox{.}(2021)]%
        {garg2021perflens}
\bibfield{author}{\bibinfo{person}{Spandan Garg}, \bibinfo{person}{Roshanak~Zilouchian Moghaddam}, \bibinfo{person}{Neel Sundaresan}, {and} \bibinfo{person}{Chen Wu}.} \bibinfo{year}{2021}\natexlab{}.
\newblock \showarticletitle{PerfLens: a data-driven performance bug detection and fix platform}. In \bibinfo{booktitle}{\emph{Proceedings of the 10th ACM SIGPLAN International Workshop on the State Of the Art in Program Analysis}}. \bibinfo{pages}{19--24}.
\newblock


\bibitem[Gautam et~al\mbox{.}(2025)]%
        {related-refactorbench}
\bibfield{author}{\bibinfo{person}{Dhruv Gautam}, \bibinfo{person}{Spandan Garg}, \bibinfo{person}{Jinu Jang}, \bibinfo{person}{Neel Sundaresan}, {and} \bibinfo{person}{Roshanak~Zilouchian Moghaddam}.} \bibinfo{year}{2025}\natexlab{}.
\newblock \bibinfo{title}{RefactorBench: Evaluating Stateful Reasoning in Language Agents Through Code}.
\newblock
\newblock
\showeprint[arxiv]{2503.07832}~[cs.AI]
\urldef\tempurl%
\url{https://arxiv.org/abs/2503.07832}
\showURL{%
\tempurl}


\bibitem[{GitHub}(2024)]%
        {copilot_agent_2024}
\bibfield{author}{\bibinfo{person}{{GitHub}}.} \bibinfo{year}{2024}\natexlab{}.
\newblock \bibinfo{title}{{GitHub Copilot Agent}}.
\newblock \bibinfo{howpublished}{\url{https://github.blog/news-insights/product-news/github-copilot-meet-the-new-coding-agent/}}.
\newblock
\newblock
\shownote{Accessed: 2025-07-14}.


\bibitem[Han et~al\mbox{.}(2012)]%
        {han-icse2012}
\bibfield{author}{\bibinfo{person}{Shi Han}, \bibinfo{person}{Yingnong Dang}, \bibinfo{person}{Song Ge}, \bibinfo{person}{Dongmei Zhang}, {and} \bibinfo{person}{Tao Xie}.} \bibinfo{year}{2012}\natexlab{}.
\newblock \showarticletitle{Performance Debugging in the Large via Mining Millions of Stack Traces}. In \bibinfo{booktitle}{\emph{Proceedings of the 34th International Conference on Software Engineering}} (Zurich, Switzerland) \emph{(\bibinfo{series}{ICSE '12})}. \bibinfo{publisher}{IEEE Press}, \bibinfo{pages}{145–155}.
\newblock
\showISBNx{9781467310673}


\bibitem[Iqbal et~al\mbox{.}(2021)]%
        {misconfigurations}
\bibfield{author}{\bibinfo{person}{Md~Shahriar Iqbal}, \bibinfo{person}{Rahul Krishna}, \bibinfo{person}{Mohammad~Ali Javidian}, \bibinfo{person}{Baishakhi Ray}, {and} \bibinfo{person}{Pooyan Jamshidi}.} \bibinfo{year}{2021}\natexlab{}.
\newblock \showarticletitle{CADET: Debugging and Fixing Misconfigurations using Counterfactual Reasoning}.
\newblock
\urldef\tempurl%
\url{https://doi.org/10.48550/arXiv.2010.06061}
\showDOI{\tempurl}


\bibitem[Jimenez et~al\mbox{.}(2024)]%
        {jimenez2024swebench}
\bibfield{author}{\bibinfo{person}{Carlos~E Jimenez}, \bibinfo{person}{John Yang}, \bibinfo{person}{Alexander Wettig}, \bibinfo{person}{Shunyu Yao}, \bibinfo{person}{Kexin Pei}, \bibinfo{person}{Ofir Press}, {and} \bibinfo{person}{Karthik~R Narasimhan}.} \bibinfo{year}{2024}\natexlab{}.
\newblock \showarticletitle{{SWE}-bench: Can Language Models Resolve Real-world Github Issues?}. In \bibinfo{booktitle}{\emph{The Twelfth International Conference on Learning Representations}}.
\newblock
\urldef\tempurl%
\url{https://openreview.net/forum?id=VTF8yNQM66}
\showURL{%
\tempurl}


\bibitem[Jovic et~al\mbox{.}(2011)]%
        {catchmeifyoucan}
\bibfield{author}{\bibinfo{person}{Milan Jovic}, \bibinfo{person}{Andrea Adamoli}, {and} \bibinfo{person}{Matthias Hauswirth}.} \bibinfo{year}{2011}\natexlab{}.
\newblock \showarticletitle{Catch Me If You Can: Performance Bug Detection in the Wild}.
\newblock \bibinfo{journal}{\emph{SIGPLAN Not.}} \bibinfo{volume}{46}, \bibinfo{number}{10} (\bibinfo{date}{oct} \bibinfo{year}{2011}), \bibinfo{pages}{155–170}.
\newblock
\showISSN{0362-1340}
\urldef\tempurl%
\url{https://doi.org/10.1145/2076021.2048081}
\showDOI{\tempurl}


\bibitem[Kim et~al\mbox{.}(2013)]%
        {auto_patch_gen}
\bibfield{author}{\bibinfo{person}{Dongsun Kim}, \bibinfo{person}{Jaechang Nam}, \bibinfo{person}{Jaewoo Song}, {and} \bibinfo{person}{Sunghun Kim}.} \bibinfo{year}{2013}\natexlab{}.
\newblock \showarticletitle{Automatic patch generation learned from human-written patches}. In \bibinfo{booktitle}{\emph{2013 35th International Conference on Software Engineering (ICSE)}}. \bibinfo{pages}{802--811}.
\newblock
\urldef\tempurl%
\url{https://doi.org/10.1109/ICSE.2013.6606626}
\showDOI{\tempurl}


\bibitem[Li et~al\mbox{.}(2024)]%
        {li2024promptinglargelanguagemodels}
\bibfield{author}{\bibinfo{person}{Bowen Li}, \bibinfo{person}{Wenhan Wu}, \bibinfo{person}{Ziwei Tang}, \bibinfo{person}{Lin Shi}, \bibinfo{person}{John Yang}, \bibinfo{person}{Jinyang Li}, \bibinfo{person}{Shunyu Yao}, \bibinfo{person}{Chen Qian}, \bibinfo{person}{Binyuan Hui}, \bibinfo{person}{Qicheng Zhang}, \bibinfo{person}{Zhiyin Yu}, \bibinfo{person}{He Du}, \bibinfo{person}{Ping Yang}, \bibinfo{person}{Dahua Lin}, \bibinfo{person}{Chao Peng}, {and} \bibinfo{person}{Kai Chen}.} \bibinfo{year}{2024}\natexlab{}.
\newblock \bibinfo{title}{Prompting Large Language Models to Tackle the Full Software Development Lifecycle: A Case Study}.
\newblock
\newblock
\showeprint[arxiv]{2403.08604}~[cs.CL]
\urldef\tempurl%
\url{https://arxiv.org/abs/2403.08604}
\showURL{%
\tempurl}


\bibitem[Liu et~al\mbox{.}(2023)]%
        {liu2023pre}
\bibfield{author}{\bibinfo{person}{Pengfei Liu}, \bibinfo{person}{Weizhe Yuan}, \bibinfo{person}{Jinlan Fu}, \bibinfo{person}{Zhengbao Jiang}, \bibinfo{person}{Hiroaki Hayashi}, {and} \bibinfo{person}{Graham Neubig}.} \bibinfo{year}{2023}\natexlab{}.
\newblock \showarticletitle{Pre-train, prompt, and predict: A systematic survey of prompting methods in natural language processing}.
\newblock \bibinfo{journal}{\emph{Comput. Surveys}} \bibinfo{volume}{55}, \bibinfo{number}{9} (\bibinfo{year}{2023}), \bibinfo{pages}{1--35}.
\newblock


\bibitem[{.NET Foundation}(2024)]%
        {benchmarkdotnet}
\bibfield{author}{\bibinfo{person}{{.NET Foundation}}.} \bibinfo{year}{2024}\natexlab{}.
\newblock \bibinfo{title}{BenchmarkDotNet}.
\newblock
\newblock
\urldef\tempurl%
\url{https://github.com/dotnet/BenchmarkDotNet}
\showURL{%
\tempurl}
\newblock
\shownote{Accessed: 2025-09-20}.


\bibitem[Nistor et~al\mbox{.}(2013)]%
        {caramelnistor}
\bibfield{author}{\bibinfo{person}{Adrian Nistor}, \bibinfo{person}{Tian Jiang}, {and} \bibinfo{person}{Lin Tan}.} \bibinfo{year}{2013}\natexlab{}.
\newblock \showarticletitle{Discovering, reporting, and fixing performance bugs}.
\newblock \bibinfo{journal}{\emph{2013 10th Working Conference on Mining Software Repositories (MSR)}} (\bibinfo{year}{2013}), \bibinfo{pages}{237--246}.
\newblock
\urldef\tempurl%
\url{https://doi.org/10.1109/MSR.2013.6624035}
\showDOI{\tempurl}


\bibitem[Song and Lu(2014)]%
        {song2014oopsla}
\bibfield{author}{\bibinfo{person}{Linhai Song} {and} \bibinfo{person}{Shan Lu}.} \bibinfo{year}{2014}\natexlab{}.
\newblock \showarticletitle{Statistical debugging for real-world performance problems}. In \bibinfo{booktitle}{\emph{Proceedings of the 2014 ACM International Conference on Object Oriented Programming Systems Languages \& Applications}}. \bibinfo{pages}{561--578}.
\newblock
\urldef\tempurl%
\url{https://doi.org/10.1145/2660193.2660234}
\showDOI{\tempurl}


\bibitem[Wang et~al\mbox{.}(2024)]%
        {wang2024opendevinopenplatformai}
\bibfield{author}{\bibinfo{person}{Xingyao Wang}, \bibinfo{person}{Boxuan Li}, \bibinfo{person}{Yufan Song}, \bibinfo{person}{Frank~F. Xu}, \bibinfo{person}{Xiangru Tang}, \bibinfo{person}{Mingchen Zhuge}, \bibinfo{person}{Jiayi Pan}, \bibinfo{person}{Yueqi Song}, \bibinfo{person}{Bowen Li}, \bibinfo{person}{Jaskirat Singh}, \bibinfo{person}{Hoang~H. Tran}, \bibinfo{person}{Fuqiang Li}, \bibinfo{person}{Ren Ma}, \bibinfo{person}{Mingzhang Zheng}, \bibinfo{person}{Bill Qian}, \bibinfo{person}{Yanjun Shao}, \bibinfo{person}{Niklas Muennighoff}, \bibinfo{person}{Yizhe Zhang}, \bibinfo{person}{Binyuan Hui}, \bibinfo{person}{Junyang Lin}, \bibinfo{person}{Robert Brennan}, \bibinfo{person}{Hao Peng}, \bibinfo{person}{Heng Ji}, {and} \bibinfo{person}{Graham Neubig}.} \bibinfo{year}{2024}\natexlab{}.
\newblock \bibinfo{title}{OpenDevin: An Open Platform for AI Software Developers as Generalist Agents}.
\newblock
\newblock
\showeprint[arxiv]{2407.16741}~[cs.SE]
\urldef\tempurl%
\url{https://arxiv.org/abs/2407.16741}
\showURL{%
\tempurl}


\bibitem[Weimer et~al\mbox{.}(2009)]%
        {genprog}
\bibfield{author}{\bibinfo{person}{Westley Weimer}, \bibinfo{person}{ThanhVu Nguyen}, \bibinfo{person}{Claire Le~Goues}, {and} \bibinfo{person}{Stephanie Forrest}.} \bibinfo{year}{2009}\natexlab{}.
\newblock \showarticletitle{Automatically finding patches using genetic programming}. In \bibinfo{booktitle}{\emph{2009 IEEE 31st International Conference on Software Engineering}}. \bibinfo{pages}{364--374}.
\newblock
\urldef\tempurl%
\url{https://doi.org/10.1109/ICSE.2009.5070536}
\showDOI{\tempurl}


\bibitem[Wen et~al\mbox{.}(2018)]%
        {capgen}
\bibfield{author}{\bibinfo{person}{Ming Wen}, \bibinfo{person}{Junjie Chen}, \bibinfo{person}{Rongxin Wu}, \bibinfo{person}{Dan Hao}, {and} \bibinfo{person}{Shing-Chi Cheung}.} \bibinfo{year}{2018}\natexlab{}.
\newblock \showarticletitle{Context-Aware Patch Generation for Better Automated Program Repair}. In \bibinfo{booktitle}{\emph{2018 IEEE/ACM 40th International Conference on Software Engineering (ICSE)}}. \bibinfo{pages}{1--11}.
\newblock
\urldef\tempurl%
\url{https://doi.org/10.1145/3180155.3180233}
\showDOI{\tempurl}


\bibitem[{Windsurf}(2024)]%
        {windsurf_2024}
\bibfield{author}{\bibinfo{person}{{Windsurf}}.} \bibinfo{year}{2024}\natexlab{}.
\newblock \bibinfo{title}{{https://windsurf.com/}}.
\newblock
\newblock
\newblock
\shownote{Accessed: 2025-07-14}.


\bibitem[Wong et~al\mbox{.}(2021)]%
        {varfix}
\bibfield{author}{\bibinfo{person}{Chu-Pan Wong}, \bibinfo{person}{Priscila Santiesteban}, \bibinfo{person}{Christian K\"{a}stner}, {and} \bibinfo{person}{Claire Le~Goues}.} \bibinfo{year}{2021}\natexlab{}.
\newblock \showarticletitle{VarFix: Balancing Edit Expressiveness and Search Effectiveness in Automated Program Repair}. In \bibinfo{booktitle}{\emph{Proceedings of the 29th ACM Joint Meeting on European Software Engineering Conference and Symposium on the Foundations of Software Engineering}} (Athens, Greece) \emph{(\bibinfo{series}{ESEC/FSE 2021})}. \bibinfo{publisher}{Association for Computing Machinery}, \bibinfo{address}{New York, NY, USA}, \bibinfo{pages}{354–366}.
\newblock
\showISBNx{9781450385626}
\urldef\tempurl%
\url{https://doi.org/10.1145/3468264.3468600}
\showDOI{\tempurl}


\bibitem[Xia et~al\mbox{.}(2024)]%
        {xia2024agentlessdemystifyingllmbasedsoftware}
\bibfield{author}{\bibinfo{person}{Chunqiu~Steven Xia}, \bibinfo{person}{Yinlin Deng}, \bibinfo{person}{Soren Dunn}, {and} \bibinfo{person}{Lingming Zhang}.} \bibinfo{year}{2024}\natexlab{}.
\newblock \bibinfo{title}{Agentless: Demystifying LLM-based Software Engineering Agents}.
\newblock
\newblock
\showeprint[arxiv]{2407.01489}~[cs.SE]
\urldef\tempurl%
\url{https://arxiv.org/abs/2407.01489}
\showURL{%
\tempurl}


\bibitem[Yang et~al\mbox{.}(2024)]%
        {yang2024sweagent}
\bibfield{author}{\bibinfo{person}{John Yang}, \bibinfo{person}{Carlos~E. Jimenez}, \bibinfo{person}{Alexander Wettig}, \bibinfo{person}{Kilian Lieret}, \bibinfo{person}{Shunyu Yao}, \bibinfo{person}{Karthik Narasimhan}, {and} \bibinfo{person}{Ofir Press}.} \bibinfo{year}{2024}\natexlab{}.
\newblock \bibinfo{title}{SWE-agent: Agent-Computer Interfaces Enable Automated Software Engineering}.
\newblock
\newblock
\showeprint[arxiv]{2405.15793}~[cs.SE]


\bibitem[Zhang et~al\mbox{.}(2024)]%
        {zhang2024autocoderover}
\bibfield{author}{\bibinfo{person}{Yuntong Zhang}, \bibinfo{person}{Haifeng Ruan}, \bibinfo{person}{Zhiyu Fan}, {and} \bibinfo{person}{Abhik Roychoudhury}.} \bibinfo{year}{2024}\natexlab{}.
\newblock \bibinfo{title}{AutoCodeRover: Autonomous Program Improvement}.
\newblock
\newblock
\showeprint[arxiv]{2404.05427}~[cs.SE]
\urldef\tempurl%
\url{https://arxiv.org/abs/2404.05427}
\showURL{%
\tempurl}


\bibitem[Zhuo et~al\mbox{.}(2025)]%
        {zhuo2025bigcodebench}
\bibfield{author}{\bibinfo{person}{Terry~Yue Zhuo}, \bibinfo{person}{Minh~Chien Vu}, \bibinfo{person}{Jenny Chim}, \bibinfo{person}{Han Hu}, \bibinfo{person}{Wenhao Yu}, \bibinfo{person}{Ratnadira Widyasari}, \bibinfo{person}{Imam Nur~Bani Yusuf}, \bibinfo{person}{Haolan Zhan}, \bibinfo{person}{Junda He}, \bibinfo{person}{Indraneil Paul}, \bibinfo{person}{Simon Brunner}, \bibinfo{person}{Chen Gong}, \bibinfo{person}{Thong Hoang}, \bibinfo{person}{Armel~Randy Zebaze}, \bibinfo{person}{Xiaoheng Hong}, \bibinfo{person}{Wen-Ding Li}, \bibinfo{person}{Jean Kaddour}, \bibinfo{person}{Ming Xu}, \bibinfo{person}{Zhihan Zhang}, \bibinfo{person}{Prateek Yadav}, \bibinfo{person}{Naman Jain}, \bibinfo{person}{Alex Gu}, \bibinfo{person}{Zhoujun Cheng}, \bibinfo{person}{Jiawei Liu}, \bibinfo{person}{Qian Liu}, \bibinfo{person}{Zijian Wang}, \bibinfo{person}{Binyuan Hui}, \bibinfo{person}{Niklas Muennighoff}, \bibinfo{person}{David Lo}, \bibinfo{person}{Daniel Fried}, \bibinfo{person}{Xiaoning Du}, \bibinfo{person}{Harm de
  Vries}, {and} \bibinfo{person}{Leandro~Von Werra}.} \bibinfo{year}{2025}\natexlab{}.
\newblock \bibinfo{title}{BigCodeBench: Benchmarking Code Generation with Diverse Function Calls and Complex Instructions}.
\newblock
\newblock
\showeprint[arxiv]{2406.15877}~[cs.SE]
\urldef\tempurl%
\url{https://arxiv.org/abs/2406.15877}
\showURL{%
\tempurl}


\end{thebibliography}

\end{document}